\documentclass[12pt]{article}
\usepackage{a4wide}
\usepackage{amssymb}
\begin{document}
{\renewcommand{\thefootnote}{\fnsymbol{footnote}}
\hfill  CGPG--02/2--2\\
\medskip
\hfill gr--qc/0202077\\
\medskip
\begin{center}
{\LARGE  Isotropic Loop Quantum Cosmology}\\
\vspace{1.5em}
Martin Bojowald\footnote{e-mail address: {\tt bojowald@gravity.phys.psu.edu}}\\
\vspace{0.5em}
Center for Gravitational Physics and Geometry,\\
The Pennsylvania State
University,\\
104 Davey Lab, University Park, PA 16802, USA\\
\vspace{1.5em}
\end{center}
}

\setcounter{footnote}{0}

\newcommand{\case}[2]{{\textstyle \frac{#1}{#2}}}
\newcommand{\lP}{l_{\mathrm P}}
\newcommand{\HE}{H^{({\mathrm E})}}
\newcommand{\hatHE}{\hat{H}^{({\mathrm E})}}

\newcommand{\md}{{\mathrm{d}}}
\newcommand{\Aut}{\mathop{\mathrm{Aut}}}
\newcommand{\Ad}{\mathop{\mathrm{Ad}}\nolimits}
\newcommand{\ad}{\mathop{\mathrm{ad}}\nolimits}
\newcommand{\Hom}{\mathop{\mathrm{Hom}}}
\newcommand{\Ima}{\mathop{\mathrm{Im}}}
\newcommand{\id}{\mathop{\mathrm{id}}}
\newcommand{\diag}{\mathop{\mathrm{diag}}}
\newcommand{\Kern}{\mathop{\mathrm{ker}}}
\newcommand{\tr}{\mathop{\mathrm{tr}}}
\newcommand{\sgn}{\mathop{\mathrm{sgn}}}
\newcommand{\dive}{\mathop{\mathrm{div}}}
\newcommand{\Diff}{\mathop{\mathrm{Diff}}\nolimits}

\newcommand*{\R}{{\mathbb R}}
\newcommand*{\N}{{\mathbb N}}
\newcommand*{\Z}{{\mathbb Z}}
\newcommand*{\Q}{{\mathbb Q}}
\newcommand*{\C}{{\mathbb C}}

\begin{abstract}
  Isotropic models in loop quantum cosmology allow explicit
  calculations, thanks largely to a completely known volume spectrum,
  which is exploited in order to write down the evolution equation in
  a discrete internal time. Because of genuinely quantum geometrical
  effects the classical singularity is absent in those models in the
  sense that the evolution does not break down there, contrary to the
  classical situation where space-time is inextendible. This effect is
  generic and does not depend on matter violating energy conditions,
  but it does depend on the factor ordering of the Hamiltonian
  constraint. Furthermore, it is shown that loop quantum cosmology
  reproduces standard quantum cosmology and hence (e.g., via WKB
  approximation) to classical behavior in the large volume regime
  where the discreteness of space is insignificant. Finally, an
  explicit solution to the Euclidean vacuum constraint is discussed
  which is the unique solution with semiclassical behavior
  representing quantum Euclidean space.
\end{abstract}

\section{Introduction}

General relativity is a very successful theory for the gravitational
field which is well tested in the weak field regime. However, it also
implies the well-known singularity theorems \cite{HawkingEllis}
according to which singularities and therefore a breakdown of this
theory are unavoidable provided that matter behaves in a classically
reasonable manner (i.e., fulfills energy conditions). In fact,
observations of the cosmic microwave background demonstrate that the
universe was in a very dense state a long time ago, which classically
can be understood only in models which have an initial singularity.

However, if the universe is small and the gravitational field strong,
the classical description is supposed to break down and to be replaced
by a quantum theory of gravity. In early attempts it was proposed that
the singularity could be avoided by coupling classical or quantum
matter fields which violate the strong energy condition and thereby
evade the singularity theorems \cite{Bahcall,Parker,Davies,Visser}.
Another approach modifies the field equations by adding higher
curvature terms interpreted as the leading order corrections of
quantum gravity \cite{Ruzmaikin,Gurovich,Starobinsky}. However, the
first approach uses matter effects rather than those of quantum
gravity, and in the second, more and more corrections are needed the
closer one comes to the classical singularity. Furthermore, the
procedure of truncating the series of perturbative corrections
(involving higher derivatives), and treating all solutions to the
resulting equations of high order on the same footing as those of the
lowest order ones, is inconsistent \cite{Simon}. Eventually one still
needs a full non-perturbative quantum theory of gravity in order to
understand the fate of the singularity.

Lacking a full quantization of general relativity, an early approach
to quantum cosmology was to perform a symmetry reduction to
homogeneous or isotropic models with a finite number of degrees of
freedom and to quantize afterwards \cite{DeWitt,Misner}; this will be
called ``standard quantum cosmology'' in the following. But in
general, those models do not avoid the classical singularity (even the
meaning of this phrase is not clearly understood: an early idea was to
impose the condition that the wave function $\psi(a)$ vanishes at the
singularity $a=0$ \cite{DeWitt}, but this is insufficient with a
continuous spectrum of the scale factor $a$ \cite{Blyth} which is
always the case in standard quantum cosmology), and more severely
quantum mechanical methods are used for the quantization which are not
believed to be applicable to a full quantization of general
relativity. Therefore, it is not clear to what extent the results are
relevant for quantum gravity.

In the meantime, a candidate for quantum gravity has emerged which is
now called quantum geometry \cite{Nonpert,Rov:Loops}. A key success of
this approach is the derivation of a discrete structure of space which
is implied by the discreteness of spectra of geometric operators like
area and volume \cite{AreaVol,Area,Vol2}. It also leads to a new
approach to quantum cosmology \cite{PhD}: By reducing this kinematical
quantum field theory to homogeneous or isotropic states using the
general framework for such a symmetry reduction at the quantum level
\cite{SymmRed} one arrives at loop quantum cosmology
\cite{cosmoI}. Due to this derivation loop quantum cosmology is very
different from standard quantum cosmology; e.g., the volume is
discrete \cite{cosmoII} which is inherited from quantum geometry. As a
consequence the Hamiltonian constraint, which governs the dynamics, of
those models \cite{cosmoIII} can be written as a difference equation
rather than the differential, Schr\"odinger- or Klein--Gordon-like
evolution equation of standard quantum cosmology; thus, also time is
discrete \cite{cosmoIV}. One may expect that the discrete structure of
space-time, which is most important at small volume, will have
dramatic consequences for the appearance of a singularity\footnote{The
author is grateful to A.\ Ashtekar for suggesting this idea and for
discussions about this issue.}. As a loose analogy one may look at the
hydrogen atom: classically it has a continuous family of orbits
leading to its instability, which is quantum mechanically cured by
allowing only a discrete set of states. Of course, one also has to
determine radiation loss or transition rates to judge stability, i.e.\
one has to take into account the fully dynamical situation.

In fact techniques developed for a quantization of the full theory
\cite{QSDI} can be used to show in isotropic models that the inverse
scale factor, whose divergence signals the singularity in the
classical model, can be quantized to a bounded operator with finite
eigenvalues even in states which are annihilated by the volume
operator \cite{InvScale}. This is already a first indication that
close to the classical singularity the system is at least better
behaved in a quantum theory of geometry, which is to be confirmed by
studying the dynamics of these models as governed by the Hamiltonian
constraint equation. The results of this analysis have been
communicated in \cite{Sing} and are described in detail in the present
paper: whereas the classical singularity presents a boundary of
space-time which can be reached in finite proper time but beyond which
an extension of space-time is impossible, the quantum evolution
equation does not break down at the classical singularity provided
only that we choose the appropriate factor ordering of the
constraint. The conclusion is, however, independent of the particular
matter coupled to the model, and so does not rely on special forms of
matter violating energy conditions. We will describe these conclusions
in detail in Section \ref{Evo} after reviewing the formalism of
isotropic loop quantum cosmology (Section \ref{Iso}) and computing the
action of the Hamiltonian constraint for spatially flat models
(Section \ref{Ham}; the constraint for positively curved models can be
found in App.\ \ref{Curved}). After discussing the fate of the
singularity in Section \ref{Fat} we will see in Section \ref{Sem}
that we simultaneously have the correct semiclassical behavior at
large volume, and present in Section \ref{Euc} an explicit solution to
the vacuum Euclidean constraint which gives quantum Euclidean space. A
comparison with other approaches and ideas put forward to avoid or
resolve the singularity can be found in Section \ref{Com} before our
conclusions where we will also discuss in which sense, compared to the
classical singularity theorems, the singularity is absent in loop
quantum cosmology.

\section{Isotropic Loop Quantum Cosmology}
\label{Iso}

A calculus for isotropic models of loop quantum cosmology can be
derived from that of quantum geometry along the lines of the general
framework for a symmetry reduction of diffeomorphism invariant quantum
field theories \cite{SymmRed}. This allows us to perform the symmetry
reduction at the quantum level by selecting symmetric states. In the
connection representation, those states are by definition only
supported on connections being invariant with respect to the given
action of the symmetry group. They can be embedded in the full
kinematical Hilbert space as distributional states. 

Specializing this framework to isotropy \cite{cosmoI} leads to
isotropic states supported on isotropic connections of the form
$A_a^i= \phi_I^i \omega_a^I= c \Lambda_I^i \omega_a^I$. Here, the
internal $SU(2)$-dreibein $\Lambda_I=\Lambda_I^i\tau_i$ is purely
gauge ($\tau_j= -\frac{i}{2} \sigma_j$ are generators of $SU(2)$ with
the Pauli matrices $\sigma_j$) and $\omega^I$ are left-invariant
one-forms on the ``translational'' part $N$ (isomorphic to $\R^3$ for
the spatially flat model or $SU(2)$ for the spatially positively
curved model) of the symmetry group $S\cong N\rtimes SO(3)$ acting on
the space manifold $\Sigma$. Orthonormality relations for the internal
dreibein are
\begin{equation}\label{ON}
 \Lambda^i_I \Lambda_i^J= \delta_I^J \quad\mbox{ and }\quad \epsilon_{ijk}
\Lambda_I^i \Lambda_J^j \Lambda_K^k= \epsilon_{IJK}\,.
\end{equation}
For homogeneous models, the nine parameters $\phi_I^i$ are arbitrary,
and for isotropic models $c$ is the only gauge-invariant parameter. A
co-triad can be expressed as $e_a^i=a_I^i \omega_a^I= a \Lambda_I^i
\omega_a^I$ with the scale factor $|a|$ (in a triad formulation it is
possible to use a variable $a$ which can take both signs even though
the two corresponding sectors are disconnected in a metric
formulation).  Using left-invariant vector fields $X_I$ fulfilling
$\omega^I(X_J)=\delta^I_J$, momenta canonically conjugate to $A_a^i$
are densitized triads of the form $E_i^a= p_i^I X_I^a= p \Lambda_i^I
X_I^a$ where $p=\sgn(a) a^2$.  Besides gauge degrees of freedom, there
are only the two canonically conjugate variables $c$ and $p$ which
have the physical meaning of curvature and square of radius, and are
coordinates of a phase space with symplectic structure
\begin{equation}\label{symp}
 \{c,p\}=\case{1}{3}\gamma\kappa
\end{equation}
($\kappa=8\pi G$ is the gravitational constant and $\gamma$ the
Barbero--Immirzi parameter; see \cite{InvScale} for an explanation of
the correct factor $\frac{1}{3}$ which is missing in \cite{cosmoII}).

Any gauge invariant isotropic state, being supported only on isotropic
connections, can be expressed as a function of $c$. An orthonormal
basis of such functions is given by the usual characters on $SU(2)$
\begin{equation}\label{chi}
 \chi_j=\frac{\sin(j+\case{1}{2})c}{\sin\case{1}{2}c}\quad,\quad
 j\in\case{1}{2}{\mathbb N}_0
\end{equation}
together with $\zeta_{-\frac{1}{2}}= (\sqrt{2} \sin\case{1}{2}c)^{-1}$
and
\begin{equation}\label{zeta}
 \zeta_j=\frac{\cos(j+\case{1}{2})c}{\sin\case{1}{2}c}\quad,\quad
 j\in\case{1}{2}{\mathbb N}_0 \,.
\end{equation}
Gauge non-invariant functions are given by $\Lambda^i_I\chi_j$ and
$\Lambda_I^i\zeta_j$ where $\Lambda_I^i$ is the internal dreibein
providing pure gauge degrees of freedom.

The states $\chi_j$, $\zeta_j$ are also eigenstates of the volume
operator \cite{cosmoII} with eigenvalues
\begin{equation}\label{vol}
 V_j=(\gamma\lP^2)^{\frac{3}{2}}\sqrt{\case{1}{27}j(j+\case{1}{2})(j+1)}\,.
\end{equation}
Because $j$ can take the value $-\frac{1}{2}$ as label of $\zeta_j$
the eigenvalue zero is threefold degenerate, whereas all other
eigenvalues are positive and twice degenerate.  An extension of the
volume operator to gauge non-invariant states is done by using the
relation $[\Lambda^i_I, \hat{V}]=0$ which follows from gauge
invariance of the volume \cite{InvScale}.

Because the basic multiplication operator is the point holonomy $h_I:=
\exp(c\Lambda_I^i\tau_i)= \cos(\frac{1}{2}c)+ 2\sin(\frac{1}{2}c)
\Lambda_I^i\tau_i$ we also need the action of $\cos\frac{1}{2}c$ and
$\sin\frac{1}{2}c$. This can be obtained in the connection
representation (\ref{chi}), (\ref{zeta}) by using trigonometric
relations, but it is easier first to introduce a new orthonormal basis
of the states
\begin{equation}
 |n\rangle:=\frac{\exp(\case{1}{2}inc)}{\sqrt{2}\sin\case{1}{2}c}
 \quad,\quad n\in\Z
\end{equation}
which are decomposed in the previous states by
\[
 |n\rangle= 2^{-\frac{1}{2}} \left(\zeta_{\frac{1}{2}(|n|-1)}+ i\sgn(n)
 \chi_{\frac{1}{2}(|n|-1)}\right)
\]
for $n\not=0$ and $|0\rangle=\zeta_{-\frac{1}{2}}$. The label $n$,
which will appear as internal time label below, is
the eigenvalue of the dreibein operator \cite{cosmoII}
\[
 \hat{p}= \widehat{\Lambda_3^iE_i^3}= -\case{1}{3}i\gamma\lP^2
 \left(\frac{\md}{\md c}+ \case{1}{2} \cot\case{1}{2}c\right)\,.
\]
On these states the action of $\cos\frac{1}{2}c$ and $\sin\frac{1}{2}c$
is
\begin{eqnarray}
 \cos(\case{1}{2}c) |n\rangle &=& \case{1}{2}\left(\exp(\case{1}{2}ic)+
   \exp(-\case{1}{2}ic)\right) \frac{\exp(\case{1}{2}inc)}{\sqrt{2}
   \sin(\case{1}{2}c)}= \case{1}{2}(|n+1\rangle+|n-1\rangle)\\
 \sin(\case{1}{2}c) |n\rangle &=& -\case{1}{2}i\left(\exp(\case{1}{2}ic)-
   \exp(-\case{1}{2}ic)\right) \frac{\exp(\case{1}{2}inc)}{\sqrt{2}
   \sin(\case{1}{2}c)}= -\case{1}{2}i(|n+1\rangle-
 |n-1\rangle)
\end{eqnarray}
and that of the volume operator is
\begin{equation}
 \hat{V}|n\rangle= V_{\frac{1}{2}(|n|-1)}|n\rangle=
 (\case{1}{6}\gamma\lP^2)^{\frac{3}{2}} \sqrt{(|n|-1)|n|(|n|+1)}
   \,|n\rangle\,.
\end{equation}

Together with the volume operator the trigonometric operators
establish a complete calculus for isotropic cosmological models and
more complicated operators can be constructed out of them.  As a first
application the expression $m_{IJ}:= q_{IJ}/\sqrt{\det q}=
|a|^{-1}\delta_{IJ}$ of the inverse scale factor has been quantized
\cite{InvScale} using techniques developed in quantum geometry in
order to quantize co-triad components, which are not fundamental
variables \cite{QSDI}. The same method, which also leads to densely
defined quantizations of matter Hamiltonians \cite{QSDV}, results in a
{\em bounded\/} operator quantizing the inverse scale factor
\begin{eqnarray}
 \hat{m}_{IJ} &=& 16(\gamma\lP^2)^{-2} \left(4 \left(\sqrt{\hat{V}}-
     \cos(\case{1}{2}c) \sqrt{\hat{V}} \cos(\case{1}{2}c) -\sin(\case{1}{2}c)
     \sqrt{\hat{V}} \sin(\case{1}{2}c)\right)^2\right. \\
  && -\delta_{IJ} \left.
   \left(\sin(\case{1}{2}c) \sqrt{\hat{V}}
     \cos(\case{1}{2}c)-\cos(\case{1}{2}c) \sqrt{\hat{V}}
   \sin(\case{1}{2}c)\right)^2\right)\nonumber
\end{eqnarray}
which has been studied in detail in \cite{InvScale}. We will later
need the action
\begin{eqnarray}\label{sVc}
 \left(\sin(\case{1}{2}c) \hat{V} \cos(\case{1}{2}c)-\cos(\case{1}{2}c) \hat{V}
   \sin(\case{1}{2}c)\right)|n\rangle &=& \case{1}{2}i \left(
   V_{\frac{1}{2}(|n+1|-1)}- V_{\frac{1}{2}(|n-1|-1)}\right) |n\rangle
 \nonumber\\
 &=& \case{1}{2}i \sgn(n)
 \left(V_{\frac{1}{2}|n|}-V_{\frac{1}{2}|n|-1}\right) |n\rangle
\end{eqnarray}
(for $n=0$ the value of $V_{-1}$ is understood to be zero).  Important
for the results of the present article is that the state
$|0\rangle=\zeta_{-\frac{1}{2}}$ is annihilated by both the volume
operator and the inverse scale factor. This would, of course, be
impossible in a classical theory and is a purely quantum geometrical
effect. (Since the singularity $a=0$ is not part of the classical
phase space, one has to extend the inverse scale factor appropriately
which is done here formally by $\sgn(a)^2/a$ \cite{InvScale}.) Our
later considerations crucially depend on the fact that all metrical
operators, and therefore all matter Hamiltonians which in some way
always contain metric components, annihilate the state $|0\rangle$.

\section{Hamiltonian Constraint for Isotropic Models}
\label{Ham}

For homogeneous models \cite{cosmoI} the Euclidean part (which is the
full constraint in Euclidean signature if $\gamma=1$) of the
Hamiltonian constraint is given by (the lapse function is irrelevant
for cosmological models and set to be one)
\begin{eqnarray}
 \HE_{\mathrm{hom}} &=& -\kappa^{-1}\det(a_I^i)^{-1}\epsilon_{ijk}
  F^i_{IJ}E^I_jE^J_k\nonumber\\
 &=& \kappa^{-1}\det(a_I^i)^{-1}
  (\epsilon_{ijk}c^K_{IJ} \phi^i_Kp^I_jp^J_k-
  \phi^j_I\phi^k_Jp^I_jp^J_k+ \phi^k_I\phi^j_Jp^I_jp^J_k)
  \label{HEhom}
\end{eqnarray}
($F_{IJ}^i$ are the curvature components of the connection $A_I^i$ and
$c^K_{IJ}$ are the structure constants of the symmetry group). Here,
we had to choose a relative sign for the constraint in the two
different orientations of the triad. Classically, both orientations
are disconnected because one has to require a non-degenerate triad.
However, this is no longer necessary in a quantum theory, and we will
in fact see that an evolution through degenerate metrics is possible.
Therefore we have to choose the relative sign which we did by using
the determinant of the co-triad instead of the metric, which would
always be positive. The quantization techniques of \cite{QSDI}
directly apply only to the convention in (\ref{HEhom}). We will see
that one can transform between the two choices after quantization,
however not unambiguously due to special features at the classical
singularity; see the remarks following (\ref{HamisoAct}).

\subsection{The Classical Constraint}

In the present context of isotropic models we are interested only in
special homogeneous models which can be further reduced to
isotropy. These are the two Bianchi class A models given by the
structure constants $c^K_{IJ}=0$ (Bianchi type I) which lead to the
isotropic spatially flat model or $c^K_{IJ}=\epsilon^K_{IJ}$ (Bianchi
type IX) leading to the isotropic model with positive spatial
curvature. The third isotropic model, which has negative spatial
curvature, can only be derived from a class B model and so is not
accessible in the present framework. Inserting isotropic connection
and triad components into (\ref{HEhom}) yields
\begin{equation}\label{HEiso}
 \HE= 6\kappa^{-1}a^{-3} (2\Gamma-c)cp^2= 6\kappa^{-1}
 (2\Gamma-c)c \sgn(p)\sqrt{|p|}
\end{equation}
where $\Gamma=0$ for the flat model and $\Gamma=\frac{1}{2}$ for the
positively curved model.

The parameter $\Gamma$ also determines the spin connection compatible
with a given triad, which is given by \cite{AsFlat}
\[
 \Gamma_a^i=-\case{1}{2}\epsilon^{ijk}e_j^b (2\partial_{[a} e_{b]}^k+
 e_k^ce_a^l \partial_c e_b^l)\,.
\]
For homogeneous triads $e_i^a=a_i^I X_I^a$ and inverse co-triads
$e_a^i= a_I^i \omega_a^I$ this specializes to
\[
 \Gamma_a^i=-\epsilon^{ijk} a_j^JX_J^b (a_K^k\partial_{[a}
 \omega_{b]}^K+ \case{1}{2} a_k^Ka_I^la_L^lX_K^c \omega_a^I
 \partial_{[c} \omega_{b]}^L)
\]
which using the Maurer--Cartan relations $\partial_{[a}\omega_{b]}^I=
-\frac{1}{2} c^I_{JK}\omega_{[a}^J \omega_{b]}^K$ yields
\[
 \Gamma_a^i= \case{1}{2} \epsilon^{ijk} (c^K_{IJ} a_j^Ja_K^k+
 \case{1}{2} c^L_{KJ} a_j^Ja_k^Ka_I^la_L^l) \omega_a^I =: \Gamma_I^i
 \omega_a^I\,.
\]
In isotropic models the co-triad has the special form $a_I^i=
a\Lambda_I^i$ which implies $a_i^I=a^{-1} \Lambda^I_i$ and leads to
\[
 \Gamma_I^i=\case{1}{2} c^K_{IJ} \epsilon^J_{KL} \Lambda^i_L+
 \case{1}{4} c^L_{KJ} \epsilon^{JKM} \Lambda_M^i \delta_{IL}
\]
using the orthonormality relations (\ref{ON}) for the internal
dreibein $\Lambda^i_I$. Now we use that for Bianchi class A models the
structure constants have the form $c^K_{IJ}= \epsilon_{IJL} n^{LK}$
where $n^{LK}= n^{(K)} \delta^{LK}$ is a diagonal matrix with
$n^{I}=0$ for Bianchi I and $n^{I}=1$ for Bianchi IX. With these
structure constants we finally arrive at
\begin{equation}
 \Gamma_I^i= \case{1}{2} (n^1+n^2+n^3-2n^{(I)}) \Lambda_I^i= \Gamma
 \Lambda_I^i
\end{equation}
where the constant $\Gamma$ which specifies the isotropic model has
been defined above. In particular, we see that the spin connection
vanishes for the spatially flat model, and so the Ashtekar connection
$A_a^i=\Gamma_a^i+\gamma K_a^i$ is proportional to the extrinsic
curvature, whereas in the positively curved model it has an extra term
given by the intrinsic curvature of space.

Knowing the spin connection we can express the integrated trace of the
extrinsic curvature in terms of the isotropic variables $(c,p)$:
\begin{eqnarray}
 K &:=& \int\md^3x K_a^iE_i^a= \gamma^{-1} \int\md^3x (A_a^i-\Gamma_a^i)
  E_i^a\nonumber\\
 &=& \gamma^{-1} (\phi_I^i-\Gamma_I^i) p_i^I= 3\gamma^{-1}
  (c-\Gamma)p=: K^i_Ip^I_i
 \label{Kiso} \,.
\end{eqnarray}
We also define the isotropic extrinsic curvature component $k$ by
$k\Lambda^i_I:= K_I^i= \gamma^{-1}(c-\Gamma)\Lambda_I^i$ such that
$K=3kp$.  As proposed in \cite{QSDI}, we will use this quantity in
order to quantize the Hamiltonian constraint for Lorentzian signature
exploiting the relation
\[
 K=\case{1}{2}\gamma^{-2}\{\HE,V\}\sgn(\det a_I^i)
\]
which can easily be verified here for the isotropic Euclidean part
(\ref{HEiso}) of the constraint with the symplectic structure
(\ref{symp}). This can then be inserted into the Lorentzian constraint
\begin{equation}
 H = -\HE+P \label{HLor}
\end{equation}
with
\begin{eqnarray}
P &:=& -2(1+\gamma^2)\kappa^{-1}\det(a_I^i)^{-1} K_{[I}^iK_{J]}^j
 E_i^IE_j^J \nonumber\\
 &=& -(1+\gamma^2)\kappa^{-1} a^{-3} (K_I^iK_J^jp_i^Ip_j^J-
 K_J^iK_I^jp_i^Ip_j^J) \nonumber\\
 &=& -6(1+\gamma^2)\kappa^{-1}a^{-3}k^2p^2=
  - 6(1+\gamma^{-2})\kappa^{-1}(c-\Gamma)^2 \sgn(p)\sqrt{|p|} \label{P}
\end{eqnarray}
to yield the Lorentzian constraint for isotropic models.

\subsection{Quantization}

According to \cite{cosmoIII}, the Hamiltonian constraint for
homogeneous models can be quantized along the lines of the full theory
\cite{QSDI} if the special requirements of the symmetry are taken into
account. One arrives at the Euclidean part (note that this corresponds
to the relative sign for the two different triad orientations as
chosen above; see the discussion following equation (\ref{HEhom}))
\begin{equation}\label{HEhomhol}
 \hatHE = 4i(\gamma\kappa\lP^2)^{-1} \sum_{IJK} \epsilon^{IJK}{\rm
 tr}(h_Ih_J h_I^{-1}h_J^{-1} h_{[I,J]}^{-1} h_K[h_K^{-1},\hat{V}])
\end{equation}
where the holonomy operator $h_{[I,J]}$ depends on the symmetry type
and is defined by
\[
 h_{[I,J]}:=\prod_K (h_K)^{c^K_{IJ}}\,.
\]
In contrast to \cite{cosmoIII} we quantized the Poisson bracket of
$\phi_I^i$ and the volume to $-h_K[h_K^{-1},\hat{V}]$ which is
completely along the lines of the full theory. Although in homogeneous
models it is possible to use the simpler expression $[h_K,\hat{V}]$,
which has been done in \cite{cosmoIII}, this is not advisable as can
be seen from a quantization of the inverse scale factor
\cite{InvScale}. Therefore, we use here the quantization which is
closer to that in the full theory.

Using the extrinsic curvature, the Lorentzian constraint operator can
be written as
\begin{eqnarray}\label{HamBianchiQuant}
 \hat{H} &=& -\hatHE+\hat{P}\nonumber\\
 &=& -\hatHE-8i(1+\gamma^2)\kappa^{-1}
  (\gamma\lP^2)^{-3} \sum_{IJK}
  \epsilon^{IJK} \tr\left(h_I[h_I^{-1},\hat{K}] h_J[h_J^{-1},\hat{K}]
   h_K[h_K^{-1},\hat{V}]\right)
\end{eqnarray}
with (up to ordering ambiguities which will be discussed below)
\begin{equation}\label{Kdef}
  \hat{K}=-\case{1}{2}i\gamma^{-2}\hbar^{-1}
  \left[\hatHE,\hat{V}\right] \widehat{\sgn(\det a_I^i)} \,.
\end{equation}
Since $\hat{K}$ appears in $\hat{P}$ within a commutator with
holonomies (corresponding to Poisson brackets in the classical
expression), the sign $\sgn(\det a_I^i)$ in (\ref{Kdef}) is important
even though $\hat{P}$ is quadratic in $\hat{K}$.

From the homogeneous operators we can always derive the isotropic ones
by inserting holonomies $h_I=\cos(\frac{1}{2}c)+2 \sin(\frac{1}{2}c)
\Lambda^i_I\tau_i$ and the isotropic volume operator. Using the
dreibein relations (\ref{ON}) for $\Lambda^i_I$ and $\tr(\Lambda_I)=0$
one can then take the trace in order to arrive at an operator composed
of $\cos(\frac{1}{2}c)$, $\sin(\frac{1}{2}c)$ and $\hat{V}$. For the
Hamiltonian constraint we need
\begin{eqnarray*}
 h_Ih_Jh_I^{-1}h_J^{-1} &=& \cos^4(\case{1}{2}c)+ 2
 (1+2\epsilon_{IJ}{}^K\Lambda_K) \sin^2(\case{1}{2}c) \cos^2(\case{1}{2}c)+
 (2\delta_{IJ}-1) \sin^4(\case{1}{2}c)\\
 && + 4(\Lambda_I-\Lambda_J)
 (1+\delta_{IJ}) \sin^3(\case{1}{2}c)\cos(\case{1}{2}c)
\end{eqnarray*}
which has been computed using $\Lambda_I\Lambda_J= \case{1}{2}
\epsilon_{IJ}{}^K\Lambda_K- \case{1}{4}\delta_{IJ}$. Similarly, we have
\begin{eqnarray*}
 h_K[h_K^{-1},\hat{V}] &=& \hat{V}-\cos(\case{1}{2}c) \hat{V}
 \cos(\case{1}{2}c)- \sin(\case{1}{2}c) \hat{V} \sin(\case{1}{2}c)\\
 &&- 2\Lambda_K \left(\sin(\case{1}{2}c) \hat{V} \cos(\case{1}{2}c)-
 \cos(\case{1}{2}c) \hat{V} \sin(\case{1}{2}c)\right)\,.
\end{eqnarray*}
These formulae are sufficient in order to derive the Euclidean part of
the Hamiltonian constraint for the spatially flat isotropic model
\begin{equation}\label{Hamiso}
 \hatHE= 96i(\gamma\kappa\lP^2)^{-1}
 \sin^2(\case{1}{2}c)\cos^2(\case{1}{2}c) \left(\sin(\case{1}{2}c)\hat{V}
 \cos(\case{1}{2}c)- \cos(\case{1}{2}c) \hat{V} \sin(\case{1}{2}c)\right)
\end{equation}
with action
\begin{equation}\label{HamisoAct}
 \hatHE|n\rangle= 3(\gamma\kappa\lP^2)^{-1}\sgn(n)
 \left(V_{\frac{1}{2}|n|}- V_{\frac{1}{2}|n|-1}\right) (|n+4\rangle
 -2|n\rangle+ |n-4\rangle)
\end{equation}
using (\ref{sVc}). Here we see that one can transform to the other
sign convention in (\ref{HEhom}) simply by dropping $\sgn(n)$. At this
place it looks unambiguous since for $n=0$ we have
$V_{\frac{1}{2}|n|}- V_{\frac{1}{2}|n|-1}=0$, but note that the
splitting into the sign and the difference of volume eigenvalues in
(\ref{sVc}) {\em is\/} ambiguous (since we use $\sgn(0):=0$, its
prefactor is not uniquely defined; in (\ref{sVc}) it has just been
extended from the general expression for positive and negative $n$).
In quantum geometry only the constraint with sign convention as in
(\ref{HEhom}) can be quantized directly and appears much more natural.

Next we can build the extrinsic curvature operator using the Euclidean
part of the Hamiltonian constraint. Since we chose a non-symmetric
ordering for the Euclidean constraint, we obtain from (\ref{Kdef}) a
non-symmetric extrinsic curvature operator
\begin{eqnarray}
 \hat{\cal K}|n\rangle &=& \case{3}{2}i \gamma^{-3}\lP^{-4}
 \left(V_{\frac{1}{2}|n|}- V_{\frac{1}{2}|n|-1}\right)\nonumber\\
 &&\times \left[
   \left(V_{\frac{1}{2}(|n+4|-1)}-V_{\frac{1}{2}(|n|-1)}\right)
   |n+4\rangle-
 \left(V_{\frac{1}{2}(|n|-1)}-V_{\frac{1}{2}(|n-4|-1)}\right)
   |n-4\rangle \right]\nonumber\\
 &=:& \case{1}{8}i\lP^2({\cal K}_n^-|n+4\rangle-
 {\cal K}_n^+|n-4\rangle)\label{Kaux}
\end{eqnarray}
where we defined the coefficients
\begin{equation}\label{Kn}
  {\cal K}_n^{\pm}:=\mp 12(\gamma\lP^2)^{-3}
  \left(V_{\frac{1}{2}|n|}- V_{\frac{1}{2}|n|-1}\right)
  \left(V_{\frac{1}{2}(|n\mp 4|-1)}- V_{\frac{1}{2}(|n|-1)}\right)
\end{equation}
which fulfill ${\cal K}_{-n}^-=-{\cal K}_n^+$ and are approximately
given by ${\cal K}_n^{\pm}\sim n$ for large $|n|$.  It is possible to
choose a symmetric ordering of $\hat{\cal K}$ without changing the
original ordering of $\hatHE$, but this is not necessary since we are
interested here only in the constraint, which need not be symmetric
(in fact, the Euclidean part {\em must not\/} be symmetric as we will
see below). Nevertheless, we will see shortly that not all orderings
for
\begin{equation}\label{K}
 \hat{K}|n\rangle := \alpha\hat{\cal K}+ (1-\alpha)\hat{\cal
 K}^{\dagger} = \case{1}{8}i\lP^2(K_n^-|n+4\rangle-
 K_n^+|n-4\rangle)
 \quad,\quad K_n^{\pm}=\alpha{\cal K}_n^{\pm}+ (1-\alpha){\cal
 K}_{n\mp 4}^{\mp}
\end{equation}
with ordering parameter $\alpha\in\R$ are allowed.  Using $\hat{K}$ in
some given ordering we obtain the potential term of the Lorentzian
constraint
\begin{eqnarray*}
 \hat{P} &=& -8i (1+\gamma^2) \kappa^{-1} (\gamma\lP^2)^{-3}
 \sum_{IJK}\epsilon^{IJK}
 \tr\left( h_I[h_I^{-1},\hat{K}] h_J[h_J^{-1},\hat{K}]
   h_K[h_K^{-1},\hat{V}] \right)\\
 &=& -96i(1+\gamma^2) \kappa^{-1} (\gamma\lP^2)^{-3}\\
 &&\times
 \left(\sin(\case{1}{2}c) \hat{K} \cos(\case{1}{2}c)-
 \cos(\case{1}{2}c) \hat{K}
 \sin(\case{1}{2}c)\right)^2 \left(\sin(\case{1}{2}c) \hat{V}
 \cos(\case{1}{2}c)- 
 \cos(\case{1}{2}c) \hat{V} \sin(\case{1}{2}c)\right)\,.
\end{eqnarray*}
We already commented on the ordering of $\hat{K}$, and it will be seen
to be crucial to order the operator containing $\hat{V}$, which
quantizes the triad components, to the right.

For $\hat{P}$ we need the action
\begin{equation}\label{sKc}
 \left(\sin(\case{1}{2}c) \hat{K} \cos(\case{1}{2}c)-
 \cos(\case{1}{2}c) \hat{K} 
 \sin(\case{1}{2}c)\right) |n\rangle 
 =: -\case{1}{8}\lP^2(k_n^-|n+4\rangle-
 k_n^+|n-4\rangle)
\end{equation}
introducing
\begin{equation}\label{kn}
 k_n^{\pm}=\case{1}{2}(K_{n+1}^{\pm}-
 K_{n-1}^{\pm})
\end{equation}
with $k_n^+=k_{-n}^-$. Inserting (\ref{sKc}), (\ref{kn}) and
(\ref{Kn}) into the action of $\hat{P}$ demonstrates that expressions
for the action of the isotropic Lorentzian constraint can be quite
cumbersome, but can be computed explicitly thanks to the completely
known volume spectrum (the equation simplifies at large volume where
the coefficients $k_n^{\pm}$ are approximately one, as used in Section
\ref{Lor}).  Fortunately, for the later discussion we will not need
the explicit action but only the crucial fact that we can order
$\hat{K}$ such that the coefficients $k_n^{\pm}$ are non-vanishing for
all $n$. In our original operator (\ref{Kaux}) the coefficients ${\cal
K}_0^{\pm}$ and ${\cal K}^{\pm}_{\pm2}$ vanish leading to
$k^{\pm}_{\pm1}=0$ which will be seen in the next section to give a
singular evolution. One can easily remedy this by ordering $\hat{K}$
symmetrically which amounts to replacing ${\cal K}_n^{\pm}$ by
$\frac{1}{2} ({\cal K}_n^{\pm}+ {\cal K}_{n\mp4}^{\mp})$ and results
in coefficients $k_n^{\pm}$ which never vanish. From now on, we use
the ordering
\[
 \hat{K}:=\case{1}{2}\left({\cal K}+ {\cal K}^{\dagger}\right)
\]
such that
\[
 K_n^{\pm}=\case{1}{2} \left({\cal K}_n^{\pm}+ {\cal K}_{n\mp
 4}^{\mp} \right)\,.
\]

Finally, we can compute the action of $\hat{P}$ by using the
previously derived operators:
\begin{eqnarray}
 \hat{P}|n\rangle &=& \case{3}{4} (1+\gamma^{-2})
 (\gamma\kappa\lP^2)^{-1} \sgn(n)
 \left(V_{\frac{1}{2}|n|}- V_{\frac{1}{2}|n|-1}\right)\label{PQuant}\\
 && \times\left(k_n^-k_{n+4}^-|n+8\rangle- (k_n^-k_{n+4}^++
   k_n^+k_{n-4}^-) |n\rangle+
   k_n^+k_{n-4}^+|n-8\rangle\right)\nonumber\,.
\end{eqnarray}

\section{Evolution}
\label{Evo}

Now we have all ingredients for the explicit form of the Hamiltonian
constraint equation and can discuss the evolution it governs. For
simplicity, we will write down the following formulae for the
spatially flat model and will only comment on possible differences in
the spatially positively curved model, but most qualitative results
apply to both cases.

\subsection{Discrete Time}

We will use the triad coefficient $p=\sgn(a) a^2$ as internal time
which classically makes sense only for positive $p$. In particular,
the evolution breaks down at the classical singularity $p=0$ so that
the two branches $p>0$ and $p<0$ are disconnected. (Note that $p$ and
$-p$ result in the same metric\footnote{The author thanks Y.~Ma for
making him aware of this issue.}, but are not identified by a gauge
transformation since the gauge group is $SO(3)$ rather than $O(3)$.
Factoring out the large gauge transformation $p\to -p$ is not allowed
since, in particular, it results in a conical singularity at $p=0$ in
the extended phase space which is used in the quantum theory.) This
setting enables us to discuss the fate of the singularity in quantum
cosmology by studying the evolution close to $p=0$. For this we need
to write the constraint equation $\hat{H}|s\rangle=0$ for a history
$|s\rangle$ as an evolution equation which can only be done in a
dreibein, rather than connection, representation since we are using a
metrical expression as internal time
\cite{KucharTime,Nonpert}. According to \cite{cosmoIV}, a dreibein
representation is defined by expanding $|s\rangle=\sum_ns_n|n\rangle$
and using the coefficients $s_n$ as a wave function in the dreibein
representation. Because of the discreteness of geometric spectra a
state in the dreibein representation is a function on a discrete set
given by $\Z$ here, and using a metrical internal time implies a
discrete time evolution \cite{cosmoIV}. With our explicit expressions
(\ref{HamisoAct}), (\ref{PQuant}) for the Hamiltonian constraint we
can write down the difference equations governing the evolution of
isotropic models.

In the dreibein representation, the Euclidean part of the constraint
acts as
\begin{eqnarray*}
 (\hatHE |s\rangle)_n &=& 3(\gamma\kappa\lP^2)^{-1} \left[\sgn(n+4)
   (V_{\frac{1}{2}|n+4|}-V_{\frac{1}{2}|n+4|-1}) s_{n+4} \right.\\
 &&- \left. 2\sgn(n)(V_{\frac{1}{2}|n|}-V_{\frac{1}{2}|n|-1}) s_{n}
   + \sgn(n-4) (V_{\frac{1}{2}|n-4|}-V_{\frac{1}{2}|n-4|-1})
   s_{n-4}\right]
\end{eqnarray*}
and $\hat{P}$ as
\begin{eqnarray*}
 (\hat{P}|s\rangle)_n &=& \case{3}{4} (1+\gamma^{-2})
 (\gamma\kappa\lP^2)^{-1} \left[\sgn(n+8)
 \left(V_{\frac{1}{2}|n+8|}- V_{\frac{1}{2}|n+8|-1}\right)
 k_{n+8}^+k_{n+4}^+s_{n+8}\right.\\
 && -\sgn(n) \left(V_{\frac{1}{2}|n|}- V_{\frac{1}{2}|n|-1}\right)
  (k_n^-k_{n+4}^++ k_n^+k_{n-4}^-) s_n\\
 && + \left.\sgn(n-8) \left(V_{\frac{1}{2}|n-8|}- V_{\frac{1}{2}|n-8|-1}\right)
    k_{n-8}^-k_{n-4}^- s_{n-8}\right]\,.
\end{eqnarray*}

In a realistic cosmological model we also need matter which enters the
evolution equation via its matter Hamiltonian, and we may include a
cosmological term. The precise form of the matter and its quantization
is not important here, and we will build it into our description by
using states $s_n(\phi)$ in the dreibein representation which are
functions of the matter degrees of freedom $\phi$. Since matter and
gravitational degrees of freedom are independent and so commute prior
to imposing the constraint, we will get a matter Hamiltonian
$\hat{H}_{\phi}$ acting on $s$ which is diagonal in the gravitational
degree of freedom $n$ for usual matter (for the rare cases of matter
with curvature couplings our discussion has to be adapted
appropriately; note that our starting point is not an effective
Hamiltonian with possible higher curvature terms which would have to
be derived only after quantization): $\hat{H}_{\phi}|n\rangle \otimes
|\phi\rangle=: |n\rangle \otimes \hat{H}_{\phi}(n) |\phi\rangle$.
Important for what follows is also that all states
$|0\rangle\otimes|\phi\rangle$ are annihilated by the matter
Hamiltonian, which is a consequence of special features of quantum
geometry.  More precisely, all terms in a matter Hamiltonian contain
metric components in order to have a scalar density which is
integrated to the Hamiltonian; and a quantization of those metric
components along \cite{QSDV} leads to an operator annihilating
$|0\rangle$, and so $\hat{H}_{\phi}(0)=0$. This is analogous to the
inverse scale factor which also annihilates $|0\rangle$ and is
possible only in quantum geometry because in a matter Hamiltonian both
the metric and the inverse metric can appear which cannot
simultaneously be zero classically (see \cite{InvScale} for a detailed
discussion).

Collecting all ingredients we arrive at the evolution equation
\begin{eqnarray}\label{Evolve}
&& \case{1}{4} (1+\gamma^{-2}) \sgn(n+8)
 \left(V_{\frac{1}{2}|n+8|}- V_{\frac{1}{2}|n+8|-1}\right)
 k_{n+8}^+k_{n+4}^+s_{n+8}(\phi)\nonumber\\
&&- \sgn(n+4)
 \left(V_{\frac{1}{2}|n+4|}- V_{\frac{1}{2}|n+4|-1}\right) s_{n+4}(\phi)
 \nonumber\\
&& -2 \sgn(n) \left(V_{\frac{1}{2}|n|}-
   V_{\frac{1}{2}|n|-1}\right) \left(\case{1}{8} (1+\gamma^{-2})
   (k_n^-k_{n+4}^++ k_n^+k_{n-4}^-) -1\right) s_n(\phi)\nonumber\\
&& - \sgn(n-4) \left(V_{\frac{1}{2}|n-4|}-
   V_{\frac{1}{2}|n-4|-1}\right) s_{n-4}(\phi) \nonumber\\
&&  + \case{1}{4} (1+\gamma^{-2}) \sgn(n-8)
  \left(V_{\frac{1}{2}|n-8|}- V_{\frac{1}{2}|n-8|-1}\right) 
    k_{n-8}^-k_{n-4}^- s_{n-8}(\phi)\nonumber\\
 &=& -\case{1}{3}\gamma\kappa\lP^2 \sgn(n)\hat{H}_{\phi}(n)s_n(\phi)
\end{eqnarray}
which, as anticipated in \cite{cosmoIV}, is a difference equation for
the coefficients $s_n(\phi)$ in the discrete label $n$ (our discrete
time) of order 16 (in \cite{cosmoIV} the order is higher since the
Poisson brackets with the volume have been quantized differently).
Note also that due to $k_n^+=k_{-n}^-$ the equation is symmetric under
time reflection $n\mapsto -n$, provided the matter Hamiltonian
fulfills $\hat{H}_{\phi}(n)=\hat{H}_{\phi}(-n)$.

The gravitational part of the evolution equation is quite complicated
and an explicit solution is possible only in simple cases like the
Euclidean vacuum equations discussed below. But the equation is
amenable to a numerical analysis because it is a difference equation,
and given some initial data we can compute subsequent components of
the wave function (in a numerical analysis one has to be aware of
possible unstable solutions \cite{Numerics}).  However, a recursive
computation is possible only as long as the highest order coefficient,
$\sgn(n+8) \left( V_{|n+8|/2}- V_{|n+8|/2-1} \right) k_{n+8}^+
k_{n+4}^+$, does not vanish. As discussed earlier, we use an ordering
of the extrinsic curvature such that $k_n^+$ never vanishes, but the
rest of the coefficient is zero if and only if $n=-8$. This means
that, starting at negative $n$, we can determine components $s_n$ of
the history $s$ only up to $n=-1$, and the coefficient for $n=0$ is
not determined by the evolution equation. Because the volume vanishes
in the state $|0\rangle$ it seems that as in the classical theory
there is a singularity in isotropic models of loop quantum cosmology
in which the evolution breaks down. This, however, is not the case as
we will show now.

\subsection{Fate of the Singularity}
\label{Fat}

We assume that we are given enough initial data for negative $n$ of
large absolute value in order to specify all initial conditions for
the difference equation of order $16$, i.e.\ we know the wave function
$s_n(\phi)$ at 16 successive times $n_0$ to $n_0+15$. From these
values we can compute all coefficients of $s_n(\phi)$ for negative
$n$, but for $n=-8$ the highest order coefficient in (\ref{Evolve})
vanishes. So instead of determining $s_0(\phi)$ the evolution equation
leads to a consistency condition for the initial data: We already now
$s_n(\phi)$ for all negative $n$ using (\ref{Evolve}) for $n\leq-9$,
and using the equation for $n=-8$ gives an additional condition
between $s_{-16}(\phi)$, $s_{-12}(\phi)$, $s_{-8}(\phi)$ and
$s_{-4}(\phi)$. Such a consistency condition is not a problem because
it serves to restrict the initial data which is welcome in order to
reduce the freedom, and it can be used in order to derive initial
conditions from the evolution equation which uniquely fix the
semiclassical branch of a wave function \cite{DynIn}. However, we now
lack an equation which would give us $s_0(\phi)$ in terms of the
initial data seemingly leading to a breakdown of the evolution as
mentioned above. But the situation is much better: We cannot determine
$s_0(\phi)$ because it drops out of the evolution equation, but it
also does not appear in the equations for $n>-8$. Therefore, we can
compute all coefficients for positive $n$ and in this sense we can
evolve through the classical singularity.  At this point we used the
crucial fact $\hat{H}_{\phi}(0)=0$, which naturally holds in
quantum geometry; otherwise, $s_0(\phi)$ would enter the evolution
equation via the matter part.

Of course, in order to determine the complete state $|s\rangle$ we
also need to know $s_0$ which, as we will show now, can be fixed
independently of the evolution. First we note that there is always a
trivial, degenerate solution to the constraint equation given by
$s_n(\phi)= \delta_{0,n} s_0(\phi)$ which is completely supported on
degenerate metrics and which is a true eigenstate of the constraint
with eigenvalue zero. In the vacuum case, it corresponds to the
classical solution $p=0$ which is of no physical interest, but with
matter there can be no classical analog because this solution arises
only due to $\hat{H}_{\phi}(0)=0$ whereas the classical matter
Hamiltonian usually diverges for $p=0$. All other solutions are
orthogonal to the degenerate solution and, therefore, must have
$s_0(\phi)=0$ demonstrating that in an evolving solution this
coefficient is fixed from the outset and we can determine the complete
state using the evolution equation (in a given solution to the
constraint equation there can be an arbitrary admixture of the
degenerate state, but it does not affect the solution at non-zero
$n$). After this discussion, one can absorb the sign factors in
(\ref{Evolve}) into the wave function, which had been done in
\cite{Sing,DynIn} for simplicity. This is free of ambiguities here for
evolving solutions where we use the condition $s_0=0$; compare the
remarks after (\ref{HamisoAct}).

We have now shown that the evolution equation (\ref{Evolve}) does not
break down at the classical singularity, and in this sense there is no
singularity in loop quantum cosmology. But in general a state will be
supported on the degenerate states $|\pm1\rangle$ in which the volume
vanishes. Although this may look problematic, the inverse scale
factor, whose classical divergence is responsible for the curvature
singularity, remains {\em finite\/} in these states. This feature of
quantum geometry, which also was very crucial in our proof of the
absence of the singularity because it implied a vanishing matter
Hamiltonian in $|0\rangle$, is the fundamental deviation from
classical geometry leading to the consequences discussed in the
present article. While we used this general property of the matter
Hamiltonian, the precise form of matter is irrelevant and so our
conclusion remains true for any standard type of matter with or
without a cosmological constant, and also for the spatially positively
curved model.

On the other hand, the factor ordering of the constraint is very
crucial, for in a different ordering the coefficient $s_0(\phi)$ would
not completely drop out of the evolution and we would not have the
degenerate solution. Ordering the triad components to the left rather
than to the right results in a coefficient $\sgn(n)(V_{|n|/2}-
V_{|n|/2-1})$ of $s_{n+k}$ replacing all $\sgn(n+k)(V_{|n+k|/2}-
V_{|n+k|/2-1})$ in (\ref{Evolve}). The highest order coefficient then
vanishes first for $n=0$ so that $s_8$ remains undetermined and will
{\em not\/} drop out of the equation for positive $n$. This results in
$s_{12}$, \ldots\ depending on $s_8$, and we no longer have a solution
similar to the completely degenerate state $s_n=s_0\delta_{n0}$.
Also, we must not use a symmetric ordering since in this case the
highest order coefficient $\sgn(n)(V_{|n|/2}-
V_{|n|/2-1})+\sgn(n+8)(V_{|n+8|/2}- V_{|n+8|/2-1})$ vanishes for
$n=-4$ and $s_4$, $s_8$, \ldots\ remain undetermined by initial data
(for different reasons not to use a symmetric ordering of the
Hamiltonian constraint in quantum general relativity see
\cite{Komar}). Thus, a non-singular evolution of the observed kind,
which is possible only for one of the three standard orderings (triads
to the left or right, and the symmetric ordering) may be used as a
criterion to fix the factor ordering ambiguity of the Hamiltonian
constraint. The ordering derived here corresponds to the one chosen in
\cite{QSDI} for the constraint in the full theory.

There are non-symmetric orderings of the constraint for which the
highest (and lowest) order coefficient never vanishes. With such an
ordering there would be no state corresponding to the classical
solution $p=0$ in the vacuum case, and a general evolving solution
would be supported on the degenerate state $|0\rangle$ with the $s_n$
depending on $s_0$. Since this state plays a special role even
kinematically (recall that quantizations of both the scale factor and
its inverse annihilate it \cite{InvScale}), a special behavior of any
evolving state, like the orthogonality to it described above, is
preferable. More importantly, a vanishing highest order coefficient
implies a consistency condition which poses initial conditions on
evolving states. As we will see in Sec.\ \ref{Euc}, the unique state
corresponding to flat Euclidean space only results with this condition
(see also \cite{DynIn} for the case with matter).  Non-vanishing
highest order coefficients are also obtained for a symmetric ordering
if one first transforms to the alternative sign convention in
(\ref{HamisoAct}). But recall that this transformation is not free of
ambiguities right at the value $n=0$ which is important for a
discussion of the singularity. Thus the choice of this ordering is
problematic. The scenario of \cite{Sing,DynIn} and the present paper
is realized in {\em only one\/} ordering of the constraint, with
triads to the right, which will always be used from now on.  This
ordering has been used previously in order to derive consistency of
the formal constraints \cite{Jacobson}. (However, the Chern--Simons
state, which has been found as a solution to the formal Euclidean
Hamiltonian constraint with a positive cosmological constant, requires
the opposite ordering \cite{Kodama2}. But since there is no physical
correspondence of this state it is not necessary to find it as a
solution, and so its disappearance does not present an argument
against an ordering; for the Lorentzian constraint it disappears,
anyway.)

Intuitively, we have the following picture of an evolving universe:
For negative times $n$ of large absolute value we start from a
classical universe with large volume. It contracts ($V_{(|n|-1)/2}$
decreases with increasing negative $n$) to reach a degenerate state of
zero volume, classically seen as a singularity, in which it bounces
off in order to enter an expanding branch and to reach again a
classical regime with large volume. The change of sign in $p$ during
the bounce means that the universe ``turns its inside out'' there
\cite{John}. A possible recollapse and an iteration of this behavior
depends on the matter content, but for the evolution close to the
singularity matter is irrelevant. What remains to show is that for
large volume we have in fact the correct semiclassical behavior, to
which we turn now.

\subsection{Semiclassical Regime}
\label{Sem}

Our evolution equation (\ref{Evolve}) is of the order 16 for the
spatially flat model and even 20 for the positively curved model, and
so there are many independent solutions (in fact, there are infinitely
many independent solutions if we take into account the matter degrees
of freedom, but we are mostly interested in the freedom coming from
the sole gravitational variable $n$). Of course, a classical
cosmological model does not have so many independent solutions, and so
most of the quantum solutions cannot have a classical counterpart. In
the present section we investigate the conditions for a solution to
have a semiclassical branch. (This issue is discussed in more detail
in \cite{DynIn}.)

We first have to define conditions for a semiclassical regime.
Obviously, the volume should be large compared to the Planck scale and
components of the curvature should be small. Moreover, continuous
space-time has to be a very good approximation to the discrete space
and time of quantum geometry. The first condition of large volume is
straightforwardly implemented by requiring $|n|$ to be large, but the
second condition for the curvature is more problematic. At this point
we have to recall that we are studying isotropic, in particular
homogeneous, models which are represented by idealized, distributional
states in the full quantum theory. Such an idealization can lead to
problems because one only has access to curvature integrated over
space rather than local curvature components. In the present context,
we may have to face infrared problems in the large volume regime
because the product of curvature and volume of space may be large even
if the local curvature is small. For instance, if we have a positive
cosmological constant $\Lambda$, it will enter the wave function in
the dimensionless combination $\Lambda p$ which diverges for
$p\to\infty$, even though the local curvature scale given by $\Lambda$
may be small.  Similarly, in the positively curved model we have the
connection coefficient $c=\frac{1}{2}+\gamma k$ where
$\Gamma=\frac{1}{2}$ comes from the spin connection and has the
meaning of the integrated intrinsic curvature of space. Therefore,
even if the extrinsic curvature $k$ is small enough, $c$ may not be
so. We will evade those infrared problems by assuming $c$ to be small
when studying the semiclassical limit. For the flat model this can
always be achieved by choosing not too large $p$.

Furthermore, contrary to a classical symmetry reduction in which a
homogeneous geometry can be slightly perturbed by adding small
non-homogeneous modes, homogeneous quantum states are distributional
and can only be approximated in the weak topology of the kinematical
Hilbert space. A consequence is the level splitting in the volume
spectrum if we break a symmetry: the simple isotropic volume spectrum
becomes increasingly complicated as in the homogeneous case or in the
full theory. In particular, whereas the isotropic volume spectrum has
an increasing level distance for large $j$, the full volume spectrum
becomes almost continuous (similarly as discussed for the area
spectrum in \cite{Area}). Such an almost continuous spectrum makes the
transition to a classical geometry with its continuous volume obvious;
but for this also a spectrum with decreasing {\em relative\/} level
distance is sufficient, as is the case in isotropic models (compare
with the equidistant energy spectrum of the harmonic oscillator which
does not prevent the correspondence to a continuous energy spectrum in
the classical regime for large energies). So compared to a given
volume, the change caused by increasing the time $n$ to $n+1$ is
always negligible and cannot be detected by a classical observer.

We incorporate this observation in our main condition ({\em
  pre-classicality} \cite{DynIn}) for a semiclassical regime: the wave
function $s_n(\phi)$ must not depend strongly on $n$ in the large
volume regime in order to be regarded as being semiclassical there.
More precisely, we demand that it is possible to interpolate between
the discrete labels $n$ and define a wave function $\psi(a):=s_{n(a)}$
with $n(a):=6\sgn(a)a^2\gamma^{-1} \lP^{-2}$ with $a$ ranging over a
continuous range (using $|a|= \sqrt{|p|}= \sqrt{\gamma} \lP
\sqrt{|n|/6}$ for large $|n|$ as interpolation points) which varies
only on scales much larger than the Planck scale.  At this point we
may have to face the above mentioned infrared problems: a cosmological
constant leads to a wave function with wave length $(\Lambda a)^{-1}$
which inevitably becomes smaller than the Planck length for large $a$.
Generically, there should be a regime in which curvatures are small
and the volume is not too large in order to allow a continuous time
approximation $\psi(a)$.

Given a wave function $\psi(a)$ interpolating the discrete function
$s_n$, we can approximate the action of the Hamiltonian constraint by
derivative operators. The basic operators are the difference operator
$\Delta=2i \sin\frac{1}{2}c$ and the mean operator
$\mu=\cos\frac{1}{2}c$ which have the leading order action
\begin{equation}\label{Delta}
 (\Delta s)_{n(a)}=s_{n(a)+1}-s_{n(a)-1}=\frac{\gamma
   \lP^2}{6a}\frac{\md\psi}{\md a} +O(\lP^5/a^5)
\end{equation}
and
\begin{equation}\label{mu}
 (\mu s)_{n(a)}=\case{1}{2}(s_{n(a)+1}+ s_{n(a)-1})= \psi(a)+
 O(\lP^4/a^4)
\end{equation}
following from a Taylor expansion. The higher order corrections also
contain higher derivatives, but in the semiclassical regime we only
need the leading order resulting in the standard Wheeler--DeWitt
operator
\[
 \kappa\hatHE\sim -96(i\Delta/2)^2\cdot\case{1}{4}a\sim -6
 \left(-\case{1}{3}i\gamma\lP^2\frac{\md}{\md(a^2)}\right)^2 a
\]
for large $a$ where we inserted $\Delta$ and $\mu\sim1$ in
(\ref{Hamiso}), used (\ref{sVc}) and expanded the volume eigenvalues
in $j\sim 3a^2\gamma^{-1}\lP^{-2}$. This is exactly what one obtains
from the classical constraint
\[
 \kappa\HE=-6c^2{\rm sgn}(p)\sqrt{|p|}
\]
in standard quantum cosmology by quantizing
$3\hat{c}=-i\gamma\lP^2\md/\md p$. In our framework, however, this is
only an approximate equation valid for large scale factors where a
continuous time approximation of the discrete time wave function is
possible. Analogously, one can show that the term $\hat{P}$ in the
Lorentzian constraint has the correct behavior semiclassically which
is also true for the spatially positively curved model. For the
Euclidean part of the constraint, this can most easily be seen in the
connection representation where the above expansion of the difference
operator corresponds to an expansion of $\sin\frac{1}{2}c$ in $c$.
Thus, by construction of the Hamiltonian constraint \cite{cosmoIII} an
expansion will result in the standard quantum cosmology expression at
leading order.  This observation also shows why small $c$ are
important in the semiclassical regime.

Arrived at the standard quantum cosmology framework, one can use
WKB-techniques in order to derive the correct semiclassical behavior.
This demonstrates, ignoring possible infrared problems, that isotropic
models with the evolution equation (\ref{Evolve}) have the correct
semiclassical behavior for large volume which is achieved in a
two-step procedure \cite{SemiClass}: First, the discrete time behavior
has to be approximated by introducing a continuous time and
interpolating the wave function which results in standard quantum
cosmology. In a second step, one can perform the classical limit in
order to arrive at the classical description. Since at fixed $\kappa$
and $\hbar$ the parameter $\gamma$ determines the scale of the
discreteness, one can also describe this by a two-fold limit
$\gamma\to 0$, $n\to\infty$ followed by $\hbar\to0$.

In this picture, a universe is fundamentally described quantum
geometrically in a discrete time, and standard quantum cosmology only
arises as an approximation which is not valid close to the
singularity. This explains the large discrepancies of standard quantum
cosmology and loop quantum cosmology regarding the fate of the
singularity. Also the issue of choosing boundary conditions, which are
usually imposed at $a=0$ in standard quantum cosmology, appears in a
different light.  In fact, our discrete time evolution can be seen to
lead to dynamical initial conditions which are derived from the
evolution equation and not imposed independently \cite{DynIn}.

Moreover, using higher order corrections in (\ref{Delta}) and
(\ref{mu}) and in the expansion of the volume eigenvalues one can
derive perturbative corrections to the standard Wheeler--DeWitt
operator. Thereby, one obtains an effective Hamiltonian containing
higher curvature and higher derivative terms. The closer one comes to
the classical singularity, the more of those perturbative corrections
are necessary, until such a description completely breaks down at the
classical singularity. Because we know the non-perturbative equation
which is discrete in time, we can see that a perturbative formulation
cannot suffice: even if one knew all perturbative corrections, it
would be very hard to see how they add up to the discrete time
behavior without knowing the non-perturbative
formulation. Furthermore, we see that a non-locality in time caused by
the discreteness is responsible for higher order corrections. This
also gives an indication as to why general relativity is
perturbatively non-renormalizable: adding local counterterms to a
local action can never result in a non-local behavior like that
observed above.

\section{Quantum Flat Space}
\label{Fla}

As we have seen, the complete Lorentzian constraint is of an awkward
form which makes it complicated to find explicit
solutions. Nevertheless, for the spatially flat model it is quite
simple at large volume, where the coefficients containing differences
of volume eigenvalues are nearly identical. The Euclidean part is then
of the form of a squared difference operator, whereas the Lorentzian
constraint is effectively of fourth order. For a complete solution we
also need to take into account the behavior at small volume, and we
will demonstrate for the Euclidean part that its values at the
classical singularity are crucial for the correspondence with the
classical situation. Recall that classically we have two solutions in
the vacuum case (Appendix \ref{EuclSpace}), a degenerate one given by
$p=0$ and Euclidean four-space characterized by $c=0$ (vanishing
extrinsic curvature). Standard quantum cosmology leads to two
independent non-degenerate solutions $\xi_1(c)=\delta(c)$ and
$\xi_2(c)=\delta'(c)$ in the connection representation only one of
which corresponds to Euclidean space. Moreover, there are problems in
standard quantum cosmology because it is not possible to invert triad
operators, even though there is no classical singularity in flat
space.

\subsection{Quantum Euclidean Space}
\label{Euc}

We have already shown that well-defined quantizations of inverse triad
operators do exist in loop quantum cosmology, so that we do not have
to deal with the second problem. The first problem of too many
solutions will now be investigated for the simplest model, the vacuum
Euclidean constraint with flat spatial slices. At first sight, it
seems to be more severe because the discrete Euclidean evolution
equation is of order eight, so we have to expect eight independent
solutions. We already know that one of them is the degenerate solution
$s_n=s_0 \delta_{n,0}$ corresponding to $p=0$, and we have to study
the remaining seven solutions which all lie in the continuous part of
the spectrum of the constraint. Since we are only interested in
solutions with classical regimes we also impose our condition for
pre-classical behavior, namely that the wave function $s_n$ does not
vary strongly from $n$ to $n+1$ for large $|n|$. This condition, which
is the only one besides the evolution equation, ensures the
possibility of semiclassical behavior for large volume and is
independent of the explicit form of classical solutions.

We have the difference equation
\begin{eqnarray}
 0 &=& \sgn(n+4) \left(V_{\frac{1}{2}|n+4|}-V_{\frac{1}{2}|n+4|-1}\right)
 s_{n+4}\nonumber
  - 2\sgn(n)\left(V_{\frac{1}{2}|n|}-V_{\frac{1}{2}|n|-1}\right)
 s_{n}\nonumber\\ 
 &&  + \sgn(n-4) \left(V_{\frac{1}{2}|n-4|}-
   V_{\frac{1}{2}|n-4|-1}\right) s_{n-4} \label{EvolveEucl}
\end{eqnarray}
which can immediately be seen to split into independent equations for
the four sequences $s_{4m}$, $s_{4m+1}$, $s_{4m+2}$, and $s_{4m+3}$
with $m\in\Z$. The first sequence contains the classical singularity
at $m=0$ and is subject to the consistency condition arising from the
vanishing highest order coefficient for $n=-4$. Therefore, it has only
one independent solution (the would-be other one is the degenerate
solution), whereas the three other sequences all have two independent
solutions because they are subject to a difference equation of order
two with never vanishing coefficients. By sticking together arbitrary
solutions for all four series we get all the seven independent
non-degenerate solutions of (\ref{EvolveEucl}). At this point, we can
already impose our selection criterion of pre-classical behavior for
large volume: In this regime, all four series have solutions
$s_{4m+i}\sim a_i(4m+i)+b_i$, $i=0\ldots3$, with an additional
consistency condition relating $a_0$ and $b_0$ which can only be
determined in the small volume regime. These solutions can be seen by
noting that for large $|n|$ the coefficients of $s_{n+4}$, $s_n$, and
$s_{n-4}$ in (\ref{EvolveEucl}) are nearly identical. Our classicality
condition then tells us that $a_0=a_1=a_2=a_3=:a$ because otherwise
the complete solution would vary strongly for large $|n|$ when jumping
between the four series: e.g., $s_{4m+1}-s_{4m}= 4(a_1-a_0)m+b_1-b_0$
becomes arbitrarily large for large $|m|$ if
$a_1-a_0\not=0$. Furthermore, the coefficients $b_i$ cannot be very
different from each other for the same reason where the difference
$b_i-b_j$ affects the amplitude but not the wave length of the
variation in the wave function. According to our condition that there
must not be a variation at the Planck scale, regardless of the
amplitude, we also have to set $b_0=b_1=b_2=b_3=:b$ (one may permit
differences in the $b_i$ bounded by some small parameter $\epsilon$,
but this will only lead to the solution with identical $b$, which is
interpreted as the semiclassical part, together with small
non-pre-classical contributions). Therefore, the classicality
condition reduces the eight parameters $a_i$, $b_i$ to only two
parameters $a$ and $b$ determining the form $s_n=an+b$ for large $|n|$
without reference to a particular classical solution. Now we are in a
position similar to that of standard loop quantum cosmology: we have
two independent solutions only one of which can correspond to
classical Euclidean four-space. But we still have the consistency
condition relating $a$ and $b$ which reduces the two independent
solutions to only one. To find this unique non-degenerate solution
with pre-classical behavior we have to use the full equation
(\ref{EvolveEucl}) also in the small volume regime and in fact right
at the classical singularity. This regime is, as demonstrated in the
preceding section, not accessible to standard quantum cosmology which
explains the fact that there are too many solutions in this approach.
However, it is not guaranteed that the unique solution corresponds to
the classical solution $c=0$ which can only be decided when we know
its explicit form. Since we need the evolution equation at the
singularity for a complete solution which also affects the large
volume behavior by fixing the relation between $a$ and $b$, we can
perform a crucial test of loop quantum cosmology for very strong
fields by comparing its unique solution with the classical solution
at large volume.

The consistency condition only appears for the sequence $s_{4m}$ on
which we can focus from now on; the remaining three sequences are then
fixed by the pre-classicality condition. We first look at the branch
for $m>0$ where we choose the only free parameter $s_4$: whereas the
coefficient $s_0=0$ is fixed (or drops out if non-zero), all other
coefficients are determined by (\ref{EvolveEucl}) which can be solved
for the highest order component
\begin{eqnarray*}
 s_n &=& \sgn(n) \left(V_{\frac{1}{2}|n|}-V_{\frac{1}{2}|n|-1}\right)^{-1}
 \left[ 2\sgn(n-4) \left(V_{\frac{1}{2}|n-4|}- V_{\frac{1}{2}|n-4|-1}
   \right) s_{n-4}\right.\\
 &&- \left.\sgn(n-8) \left(V_{\frac{1}{2}|n-8|}-
     V_{\frac{1}{2}|n-8|-1} \right) s_{n-8}\right]
\end{eqnarray*}
when $n\not=0$. For positive $n=4m$, $m\in\N$, we obtain successively
\begin{eqnarray*}
 s_8 &=& 2\frac{V_2-V_1}{V_4-V_3}s_4\,,\\
 s_{12} &=& \frac{2(V_4-V_3)s_8- (V_2-V_1)s_4}{V_6-V_5}=
 3\frac{V_2-V_1}{V_6-V_5} s_4
\end{eqnarray*}
and so on, leading by induction to
\begin{equation}\label{sol}
 s_{4m}=|m|\frac{V_2-V_1}{V_{2|m|}-V_{2|m|-1}}s_4\,.
\end{equation}
For negative $n$ we can do the same, leading to $s_{-4}$,
$s_{-8}$, \ldots\ in terms of $s_4$. The result is (\ref{sol}), now
with arbitrary $m\in\Z$, which is the exact solution for the sequence
$s_{4m}$ and automatically fulfills the consistency condition since we
started from $s_0=0$. Implicitly, (\ref{sol}) determines the relation
between $a$ and $b$ such that we now also have the unique solution
with pre-classical behavior given by
\begin{equation}
 s_{n}=\case{1}{4}|n|\frac{V_2-V_1}{V_{\frac{1}{2}|n|}-
   V_{\frac{1}{2}|n|-1}} s_4\,.
\end{equation}
Dropping constant factors (or choosing $s_4:=2(V_2-V_1)^{-1}$), this
yields in the connection representation
\begin{equation}\label{solj}
 \psi(c)=\sum_{n\in\Z}s_n|n\rangle= \sum_{j=0}^{\infty}
 \frac{2j+1}{V_{j+\frac{1}{2}}- V_{j-\frac{1}{2}}} \zeta_j(c)
\end{equation}
where the expression $2j+1=|n|$ (rather than another linear function
in $j$) appears because the consistency condition fixed the relation
between $a$ and $b$. This allows us to check the compatibility of this
condition, which arose because of the structure at the classical
singularity, with the correct classical behavior: we expect a solution
which is related to the $\delta$-function in $c$ incorporating the
classical solution $c=0$ (see also App.\ \ref{EuclSpace}).

More precisely, since we chose an ordering with the triad components
to the right in the connection representation, which has been seen to
be necessary in order to remove the singularity, we should expect a
quantization of $a=\sgn(p)\sqrt{|p|}$ to map $\psi$ to the delta
function (the classical constraint is $-6c^2\sgn(p)\sqrt{|p|}$ in the
ordering chosen for the quantization). A possible quantization of $a$
maps
\[
 \hat{a}\zeta_j=
 2i(\gamma\lP^2)^{-1}(V_{j+\frac{1}{2}}-V_{j-\frac{1}{2}}) \chi_j
\]
with an analogous action on $\chi_j$, which has the correct large-$j$
behavior for $\sqrt{|p|}$ and which maps $\zeta_j$ to $i\chi_j$ (and
to zero for $j=-\frac{1}{2}$). The last property is necessary due to
the $\sgn(p)$ which can be seen from the action on $|n\rangle$. One
can also use the techniques of \cite{InvScale} in order to derive this
quantization: writing
\[
 \sgn(p)\sqrt{|p|}=a =\case{1}{3}\Lambda_i^Ia_I^i= -\case{2}{3}
 \tr(\Lambda^I a_I^i\tau_i)
\]
and using Thiemann's quantization of the co-triad components we have
\[
 \hat{a}=-\case{4}{3}i(\gamma\lP^2)^{-1}\sum_I \tr(\Lambda^I h_I[h_I^{-1},
 \hat{V}])= -4i(\gamma\lP^2)^{-1} \left(\sin(\case{1}{2}c)\hat{V}
 \cos(\case{1}{2}c)- \cos(\case{1}{2}c) \hat{V} \sin(\case{1}{2}c)\right)
\]
acting as
\begin{eqnarray}
 \hat{a}\chi_j &=& -2i(\gamma\lP^2)^{-1} (V_{j+\frac{1}{2}}-
 V_{j-\frac{1}{2}}) \zeta_j\nonumber\\
 \hat{a}\zeta_j &=& 2i(\gamma\lP^2)^{-1} (V_{j+\frac{1}{2}}-
 V_{j-\frac{1}{2}}) \chi_j\,. \label{azeta}
\end{eqnarray}
The eigenvalues of $\hat{a}$ are
\[
 a_j=\pm2(\gamma\lP^2)^{-1} (V_{j+\frac{1}{2}}- V_{j-\frac{1}{2}})
\]
which for large $j$ is
\[
 |a_j|\sim\sqrt{\case{1}{3}\gamma\lP^2j}\sim V_j^{\frac{1}{3}}\,.
\]

Applying $\hat{a}$ of (\ref{azeta}) to our solution (\ref{solj})
yields
\[
 \hat{a}\psi(c)=2i(\gamma\lP^2)^{-1}\sum_j(2j+1)\chi_j(c)=
 2i(\gamma\lP^2)^{-1} \delta(c)
\]
which in fact is proportional to the $\delta$-function on the
configuration space $SU(2)$. For this it is crucial that we have the
factor $2j+1$ which appears uniquely only if one uses the consistency
condition arising from the behavior at the classical singularity.
Thus, the unique solution to the Euclidean constraint with
semiclassical behavior in spatially flat isotropic loop quantum
cosmology correctly corresponds to Euclidean four-space with vanishing
extrinsic curvature. Similarly in other models, the consistency
condition together with the pre-classicality condition on the
variation of the wave function for large $a$ always selects a unique
solution. In this way {\em dynamical initial conditions\/} are derived
from the evolution equation \cite{DynIn} and not imposed ad hoc as
usually done in standard quantum cosmology.

\subsection{Lorentzian Constraint at Large Volume}
\label{Lor}

For large $|n|$ the Lorentzian constraint equation (\ref{Evolve})
simplifies since $V_{|n+k|/2}-V_{(|n+k|-1)/2}$ is approximately
independent of $k$ for $n\gg k$ and the extrinsic curvature
coefficients $k_n^{\pm}$ are nearly one. Introducing $t_m:=s_{2m}$,
the vacuum constraint equation takes the form
\begin{eqnarray*}
 0 &=& \case{1}{4}(1+\gamma^{-2}) t_{m+4}- t_{m+2}+ \case{1}{2}
  (3-\gamma^{-2}) t_{m}- t_{m-2}+ \case{1}{4}(1+\gamma^{-2}) t_{m-4}\\
  &=& \case{1}{4}[(\Delta^4+ 4\gamma^{-2} \Delta^2 \mu^2)t]_m\\
  &=& \case{1}{4}[\Delta^2(\Delta+ 2i\gamma^{-1} \mu) (\Delta-
  2i\gamma^{-1} \mu)t]_m
\end{eqnarray*}
using again the central difference and mean operators $\Delta$ and
$\mu$. Because they commute we can split this equation into
\begin{eqnarray*}
  \Delta^2t &=& 0\\
  \mbox{or}\quad(\Delta+2i\gamma^{-1}\mu)t &=& 0\\
  \mbox{or}\quad(\Delta-2i\gamma^{-1}\mu)t &=& 0
\end{eqnarray*}
with independent solutions
\begin{eqnarray*}
 &&t^{(1)}_m=c_1\quad,\quad t^{(2)}_m=c_2 m\\
 &&t^{(3)}_m=c_3
 \left(\frac{1-i\gamma^{-1}}{1+i\gamma^{-1}}\right)^m \quad,\quad
 t^{(4)}_m= c_4 \left(\frac{1-i\gamma^{-1}}{1+i\gamma^{-1}}\right)^{-m}\,.
\end{eqnarray*}
For $\gamma=1$ we can also use the form
\[
 t^{(3)}_m=c_3 \,\mbox{Re}(i^m)=\cos(m\case{\pi}{2}) \quad,\quad
 t^{(4)}_m= c_4 \,\mbox{Im}(i^m)=\sin(m\case{\pi}{2})\,.
\]

As with the Euclidean constraint we have still more solutions since
the original equation (\ref{Evolve}) is of order 16. But again it
splits into four sequences with solutions as above which can be put
together to form all 16 independent solutions. Thanks to the
pre-classicality condition of mild variation only four of them are
relevant. However, only two, $t^{(1)}$ and $t^{(2)}$, can correspond
to the two solutions of the second order standard Wheeler--DeWitt
equation; the rest has to be excluded on general grounds which again
are provided by pre-classicality. This immediately excludes $t^{(3)}$
and $t^{(4)}$ if $\gamma=1$ which jump between the values $0$ and
$\pm1$, and also for $\gamma\not=1$ if $\gamma$ is of order one or
less (which is true for the physical value
\cite{ABCK:LoopEntro,IHEntro}): $t_m^{(3)}$ and $t_m^{(4)}$ are of the
form $\exp(\pm im\theta)$ with $\theta= \arccos [(1-\gamma^{-2})/
(1+\gamma^{-2})]$ which is $\pi>\theta>\frac{\pi}{2}$ for
$0<\gamma<1$.

We now arrived at two independent solutions allowed for a
semiclassical analysis from which a particular combination is selected
by the consistency condition at the classical singularity. To evaluate
this we would need to take into account the exact equation
(\ref{Evolve}) also for small volume, from which we refrain here. This
appears not to be possible analytically in closed form, but can easily
be done in a numerical study. Nevertheless, it is easy to see that the
Hamiltonian constraint equation has solutions of the same
semiclassical behavior for all values of $\gamma$, which contradicts
the hope \cite{Immirzi} that the Hamiltonian constraint equation
selects a value for $\gamma$. We can only conclude that a large
parameter $\gamma\gg 1$ would lead to additional pre-classical
solutions lacking any correspondence to a classical solution (e.g.,
oscillating solutions for flat space). This may be taken as an
argument that $\gamma$ cannot be much larger than one, in coincidence
with \cite{ABCK:LoopEntro,IHEntro}.

\section{Comparison with Other Approaches}
\label{Com}

A resolution or avoidance of the classical singularity has been
claimed before in a variety of approaches. After the singularity
theorems of general relativity \cite{HawkingEllis} had been
established, it became clear that one has to couple matter which
violates energy conditions in order to evade them (it is sufficient to
violate only the strong energy condition \cite{Visser}). This can be
achieved with either classical \cite{Bahcall} or quantum matter
\cite{Parker,Davies} leading to a bounce in the evolution of a
universe at positive radius. However, this conclusion is model and
parameter dependent and, therefore, not a generic
behavior. Furthermore, the threat of a singularity is still present in
gravity, but only avoided by particular types of matter (and, since
quantum matter field theories have their own divergences, it may be
dangerous to call upon quantum matter for a rescue from the
singularity).

Another idea to evade the singularity theorems consists in changing
general relativity. Since its action is deemed to be only an effective
action of something more fundamental, there can be correction terms
being non-linear in the Ricci scalar $R$. Inclusion of the lowest
order correction quadratic in $R$ has been shown to yield solutions
which do not encounter a singularity
\cite{Ruzmaikin,Gurovich,Starobinsky}. Again, the conclusion is
parameter dependent and, in fact, inconsistent because the
non-singular solutions emerge only as artifacts of the truncation of
the higher order corrections \cite{Simon}. This situation suggests
that a complete non-perturbative formulation is necessary for an
investigation of the fate of the classical singularity.

The original approach to this problem in canonical quantum gravity,
which is non-perturbative, was started in \cite{DeWitt}. Here, it was
proposed that one should use the boundary condition $\psi(0)=0$ of a
vanishing wave function right at the classical singularity. It has
also been speculated that this condition is enforced by an ad hoc
Planck potential which is relevant only for small scale factors
\cite{SIC}. However, as argued in \cite{Blyth}, this requirement by
itself cannot be regarded as a sufficient condition for the absence of
a singularity because the scale factor has a continuous spectrum in
standard quantum cosmology (note that in loop quantum cosmology the
spectrum of the scale factor is discrete, but the mechanism which
removes the classical singularity is very different from DeWitt's
proposal); in addition an appropriate fall-off behavior of $\psi$
close to $a=0$ would be necessary. In this context, it has also been
suggested, sometimes for purely mathematical reasons in order to
obtain a selfadjoint $i\md/\md p$, to extend minisuperspace to include
negative volumes, which would remove the boundary at $a=0$ and allow
wave functions to extend into this regime \cite{Kodama2} (although
this may seem similar to the evolution through the singularity derived
in the present paper, it is not to be confused with our negative $p$
which still leads to {\em positive\/} volume: in contrast to negative
definite metrics, negative triads are allowed classically even though
disconnected from the $p>0$ sector if one requires
non-degeneracy). The negative metric branch lacks a classical
interpretation, and so the wave function is completely quantum without
semiclassical interpretation in this large region of the configuration
space. It has been suggested that the transition to negative volume
should be interpreted as a signature change to Euclidean space-time
\cite{nobound} or a ``tunneling from nothing'' \cite{tunneling}.

As derived in this paper, loop quantum cosmology is able to describe
the behavior of a universe close to the classical singularity. It
always leads to a decoupling of $s_0$ from solutions of cosmological
behavior, reminiscent of DeWitt's $\psi(0)=0$. However, since from the
present perspective standard quantum cosmology completely breaks down
in this regime, a condition for the standard wave function $\psi(a)$
is no longer meaningful at $a=0$. This condition for $s_0$ is in fact
important for the absence of a singularity, and it serves to select a
unique superposition of the two WKB components when evolved into the
semiclassical regime \cite{DynIn}. At $a=0$ we have a transition to
another branch which opens to large {\em positive\/} volume at {\em
negative\/} time (rather than a ``tunneling from nothing'', our
picture of an evolving universe could be described as ``tunneling
through nothing'' if one wants to identify ``nothing'' with the
degenerate state $|0\rangle$).

This picture of a branch preceding the classical singularity is
reminiscent of the ``pre-big-bang'' scenario (and other, more recent
constructions) of string cosmology
\cite{Veneziano,VenezianoLesHouches} where a contracting universe
preceding the singularity has been claimed which should be connected
to the present expanding branch through a high curvature regime.
However, lacking a non-perturbative framework, this claim cannot be
substantiated at present.

The possibility of an evolution through a degenerate state also
reminds of results which have been obtained in the context of mirror
symmetry: by mapping the degenerate state to an equivalent
non-degenerate one it is possible to extend the evolution through a
singular geometry \cite{Mirror,Trans}. However, this has been
demonstrated only for very special spaces where one generally focuses
on the geometry of compact directions; these spaces are not
sufficiently realistic to be of direct physical interest, let alone
cosmological models.

\section{Conclusions}
\label{Con}

A reduction of quantum geometry to isotropic geometries leads to
models in which explicit computations are possible, thanks primarily
to the completely known volume spectrum. Therefore, they provide an
ideal test arena for the techniques of quantum geometry, which turn
out to work without any problems. Moreover, they are interesting in
their own right as cosmological models where they shed light on
certain aspects of the classical singularity which were not
illuminated in any other approach to (quantum) cosmology. A
kinematical indication for a better behavior of loop quantum cosmology
close to the classical singularity comes from a quantization of the
inverse scale factor \cite{InvScale} using techniques for the
quantization of matter Hamiltonians to densely defined operators in
quantum geometry \cite{QSDV}. The result is a bounded inverse scale
factor which does not diverge even if the volume is zero, a result
which is possible only in a quantum theory of geometry. In the present
paper we established the absence of a singularity by studying
evolution equations at the dynamical level: whereas classically the
evolution breaks down at the singularity, in loop quantum cosmology we
can evolve through it. Since evolution through a degenerate state is
possible, one could also obtain topology change in quantum geometry.

At this point we explain in more detail in which sense the singularity
is absent in loop quantum cosmology. One might think that there is
still a singular space geometry of vanishing volume. But this is not
as problematic as in the classical theory since, e.g., the inverse
scale factor does not diverge. One should also keep in mind that
vanishing of volume is possible even classically without a
singularity: it may just signal the presence of a horizon as is the
case in (non-singular) de Sitter space-time when sliced by flat
spaces. In this case one can, of course, evolve through the horizon by
choosing an appropriate time coordinate. In contrast to a singular
space-time, such a manifold is not geodesically incomplete. On the
other hand, the existence of an incomplete curve together with energy
conditions for the matter inevitably leads to a curvature singularity
\cite{HawkingEllis}. At such a point, the curvature tensor cannot even
be interpreted in a distributional sense and so Einstein's field
equations break down. There is then no means to extend the singular
space-time beyond the singularity in a unique manner (it may be
possible to extend a space-time continuously, but never uniquely). The
evolution equation (\ref{Evolve}) of isotropic loop quantum cosmology,
on the contrary, never breaks down and so always gives rise to a
unique extension through the quantum regime containing the classical
singularity.

In deriving this behavior it was our strategy always to be as close to
the full theory as possible, e.g.\ when quantizing the Hamiltonian
constraint. Although one might have simplified some expressions in a
model-dependent way, this would have lead to deviations from the
methods of the full theory. For instance in the case of flat spatial
slices the classical Lorentzian constraint and its Euclidean part only
differ by a factor $\gamma^{-2}$ so that one might be tempted to use
this relation also in a quantization which would strongly simplify the
quantum constraint. However, this fact crucially depends on i)
$K_a^i\propto A_a^i$ (due to $\Gamma_a^i=0$) and ii)
$\epsilon_{ijk}A_a^iA_b^j\propto F_{ab}^k$ (due to $\md A=0$) both of
which fail in other homogeneous models, let alone in the full
theory. Making use of these relations would simplify the computation,
but the results were not trustworthy since the contact to the full
theory would have been lost. Note that in standard quantum cosmology
one does not have a corresponding quantization of the full theory as
guidance, and thus lacks means to evaluate manipulations. In fact,
loop and standard quantum cosmology differ from each other right in
the regime where quantum gravitational effects are important. Standard
quantum cosmology uses quantum mechanical methods and phenomena (like
the tunneling effect) but not full quantum gravity, whereas loop
quantum gravity is close to the full theory of quantum geometry. Some
of its quantization techniques may seem to be unfamiliar from the
viewpoint of quantum mechanics, but they are necessary since analogous
techniques are required for a consistent quantization of
gravity. Moreover, quantization ambiguities (like the one described at
the beginning of this paragraph) are severely restricted by requiring
that analogous quantizations must be possible in the full theory.

For the absence of a singularity the form of the matter coupled to
gravity is irrelevant because the removal of the singularity is
completely due to quantum geometry. On the other hand, the
(non-symmetric) factor ordering of the Hamiltonian constraint is
crucial for this result, so that demanding a non-singular evolution in
quantum cosmology fixes the factor ordering ambiguity of the
constraint: the scenario derived in this paper is possible with only
one ordering, which belongs to the three standard choices (one could
still choose different orderings of the extrinsic curvature operator
entering the Lorentzian constraint, but since it is an observable in
the kinematical sector it should be ordered symmetrically as done here).

Close to the classical singularity the discrete structure of space and
time in quantum geometry is important which leads to large deviations
from standard quantum cosmology \cite{DeWitt}. This framework arises
here as an approximation which is good only at large volume where the
discrete volume spectrum is washed out to a continuous spectrum by
inaccuracies. However, the exact description of loop quantum cosmology
is also necessary to fix a unique solution which can be seen in the
explicit solution corresponding to Euclidean four-space. For
cosmological models with matter there is still a unique solution with
appropriate semiclassical behavior: initial conditions are not imposed
ad hoc but instead {\em derived\/} from the dynamical laws
\cite{DynIn}. Taking into account the discreteness of the spectrum one
can derive perturbative corrections for an effective Hamiltonian of
standard quantum cosmology, but the completely non-perturbative
description with discrete time is needed in order to study the fate of
the classical singularity.

As an intuitive picture of the evolution of a universe in loop quantum
cosmology we obtain the following one: starting in a semiclassical
contracting state, it reaches a degenerate stage seen as the
singularity in classical cosmology, in which the universe bounces off
in order to enter an expanding branch. The further fate, whether it
expands forever or recollapses in order to start a new such process,
depends on the matter content and the value of the cosmological
constant.

\vspace{1cm}

\begin{appendix}
\renewcommand{\theequation}{\thesection.\arabic{equation}}
\setcounter{equation}{0}

\section*{Appendices}
\section{Euclidean Space in Standard Quantum Cosmology}
\label{EuclSpace}

Here we present classical and standard quantum cosmology of flat
Euclidean space. For $\Gamma=0$ the constraint equation (\ref{HEiso})
becomes $\HE=-6\kappa^{-1}c^2\sgn(p)\sqrt{|p|}=0$ having two
solutions $p=0$ or $c=0$. The first one only appears in a triad
formulation and is of no physical interest because space is completely
degenerate. Using (\ref{Kiso}) one can see that the second solution
requires the extrinsic curvature to vanish, whereas it leaves the
metric arbitrary, and thus gives flat Euclidean space (the metric does
not change in coordinate time because of
$\{p,\HE\}\approx0$ on the constraint surface). 

Standard quantum cosmology proceeds as follows: Using a factor
ordering in which the metric variables appear on the right (this is
the ordering resulting from loop quantum cosmology) solutions in the
$c$-representation are such that a quantization of $\sgn(p)\sqrt{|p|}$
acting on the wave function yields either $\delta(c)$ or $\delta'(c)$
due to the factor $c^2$. There are some problems already in this
simple model: First, $\sgn(p)\sqrt{|p|}$ cannot be quantized to an
invertible operator when the range of $p$ contains the value zero, and
a procedure like that leading to a bounded inverse scale factor in
loop quantum cosmology is not available in standard quantum
cosmology. Although Euclidean space does not contain a singularity,
the point $p=0$ leads to problems in the quantization. Second, there
are two independent solutions only one of which, $\delta(c)$,
corresponds to the classical solution; and there is no independent
argument to exclude the other solution without referring explicitly to
the classical situation. Both problems are solved in loop quantum
cosmology, which is intimately related to the fate of the classical
singularity.

\section{Hamiltonian Constraint for Models with Positive Spatial
Curvature}
\label{Curved}

For the isotropic, spatially positively curved model we need to take
into account the holonomy $h_{[I,J]}$ in (\ref{HEhomhol}) which is $h_K$ if
$\epsilon_{IJK}$ is positive and $h_K^{-1}$ if $\epsilon_{IJK}$ is
negative. With this we have for isotropic holonomies
\begin{eqnarray*}
 \sum_{IJK}\epsilon^{IJK} \tr\left(h_Ih_Jh_I^{-1}h_J^{-1}
   h_{[I,J]}^{-1} h_K[h_K^{-1},\hat{V}]\right) &=& 3\tr\left(
   h_1h_2h_1^{-1}h_2^{-1} [h_3^{-1},\hat{V}]\right)\\
 &&- 3\tr\left(
   h_2h_1h_2^{-1}h_1^{-1} h_3^2 [h_3^{-1},\hat{V}]\right)
\end{eqnarray*}
which yields
\begin{eqnarray*}
 \hatHE_+ &=& -48i (\gamma\kappa\lP^2)^{-1} \left(\sin(\case{1}{2}c)-
 2\sin^5(\case{1}{2}c)-
 2\sin^2(\case{1}{2}c)\cos^3(\case{1}{2}c)\right)\\
 &&\times
 \left(\sin(\case{1}{2}c) \hat{V} \cos(\case{1}{2}c)-
 \cos(\case{1}{2}c) \hat{V} \sin(\case{1}{2}c)\right)
\end{eqnarray*}
with action
\begin{eqnarray*}
  \hatHE_+|n\rangle &=& 3(\gamma\kappa\lP^2)^{-1}\sgn(n)
  \left(V_{\frac{1}{2}|n|}-V_{\frac{1}{2}|n|-1}\right) \left(\case{1}{2}
  (1+i) |n+5\rangle+\case{1}{2}(1-5i)|n+3\rangle
  \right.\\
 && -\left.(1-i)|n+1\rangle- (1+i)|n-1\rangle+ \case{1}{2}(1+5i)|n-3\rangle+
  \case{1}{2}(1-i)|n-5\rangle\right)\,.
\end{eqnarray*}

From this we obtain the extrinsic curvature operator
\[
 \hat{K}_+|n\rangle= \case{1}{8}\lP^2 \sum_{q=-5\, ; \, q \mbox{
 \footnotesize odd}}^5 K_{+,n}^{(-q)}|n+q\rangle
\]
with
\begin{eqnarray*}
 K_{+,n}^{(\pm1)} &=& \mp 6(1\mp i)(\gamma\lP^2)^{-3}
 \left(V_{\frac{1}{2}|n|}- V_{\frac{1}{2}|n|-1}\right)
 \left(V_{\frac{1}{2}(|n\mp 1|-1)}- V_{\frac{1}{2}(|n|-1)}\right) \\
 K_{+,n}^{(\pm3)} &=& \pm 6(5\mp i)(\gamma\lP^2)^{-3}
 \left(V_{\frac{1}{2}|n|}- V_{\frac{1}{2}|n|-1}\right)
 \left(V_{\frac{1}{2}(|n\mp 3|-1)}- V_{\frac{1}{2}(|n|-1)}\right) \\
 K_{+,n}^{(\pm5)} &=& \mp 6(1\pm i)(\gamma\lP^2)^{-3}
 \left(V_{\frac{1}{2}|n|}- V_{\frac{1}{2}|n|-1}\right)
 \left(V_{\frac{1}{2}(|n\mp 5|-1)}- V_{\frac{1}{2}(|n|-1)}\right)\,.
\end{eqnarray*}
This leads to
\[
 \left(\sin(\case{1}{2}c)\hat{K}_+\cos(\case{1}{2}c)-
 \cos(\case{1}{2}c)\hat{K}_+\sin(\case{1}{2}c)\right)|n\rangle=
 \case{1}{8}i\lP^2 
 \sum_{q=-5\, ;\, q\mbox{ \footnotesize odd}}^5
 k_{+,n}^{(-q)}|n+q\rangle
\]
with
\[
  k_{+,n}^{(q)}=\case{1}{2}(K_{+,n+1}^{(q)}-K_{+,n-1}^{(q)})
\]
which is non-zero for all $n$ and $q$ ($K_{+,n}^{(q)}$ is zero if and
only if $n=0$).

Taken together, this yields
\[
 \hat{P}_+|n\rangle= -\case{3}{4}(1+\gamma^{-2})
 (\gamma\kappa\lP^2)^{-1} \sgn(n) \left(V_{\frac{1}{2}|n|}-
 V_{\frac{1}{2}|n|-1}\right) \sum_{k=-5}^5 A_n^{(-2k)} |n+2k\rangle
\]
with
\[
 A_n^{(l)}:= \sum_{q+r=l\, ;\, -5\leq q,r\leq 5 \, ;\, q,r\mbox{
 \footnotesize odd}}
 k_{+,n}^{(q)} k_{+,n-q}^{(r)}\,.
\]

\end{appendix}

\section*{Acknowledgements}

The author is grateful to A.~Ashtekar for suggestions, discussions,
and a careful reading of the manuscript.
This work was supported in part by NSF grant PHY00-90091 and the Eberly
research funds of Penn State.

\end{document}